\documentclass[twocolumn,secnumarabic,amssymb, nobibnotes, aps, prd]{revtex4-2}

\usepackage{graphicx}
\usepackage{amsmath}  
\usepackage{bm} 
\usepackage{hyperref}
\usepackage{afterpage} 
\usepackage{comment}
\usepackage{balance}  
\usepackage{amssymb}
\usepackage{booktabs} 

\begin{document}

\title{{\Large \textbf{Uncovering the maximum chirality in dielectric nanostructures}}}%
\author{Wen Kui Zhao\textsuperscript{1,{\hyperlink{note1}{†}}}}
\author{ShengYi Wang\textsuperscript{1,2,{\hyperlink{note2}{†}}}}
\author{Hanzhuo Kuang,\textsuperscript{1} Qiu Wang,\textsuperscript{1} Bo-Wen Jia\textsuperscript{1,}}%
\email[]{jiabowen@whut.edu.cn}
\affiliation{
\textsuperscript{1}School of Information Engineering, Wuhan University of Technology, Wuhan 430070, China}
\affiliation{\textsuperscript{2}Key Laboratory of Artificial Micro- and Nano-structures of Ministry of Education, and School of Physical and Technology, Wuhan University, Wuhan 430072, China}
\affiliation{\textsuperscript{3}School of Physics, Huazhong University of Science and Technology, Luoyu Road 1037, Wuhan, 430074, China}
\date{December 18 2024}%

\begin{abstract}

  Maximum structural chirality refers to the highest selectivity for circularly polarized light (CPL) in nanostructures, often manifested as maximum circular dichroism (CD), optical rotation (OR), and spin-orbit coupling (SOC). However, the underlying physical mechanisms that lead to  maximum chirality remain unclear. In this work, we demonstrate that maximum chirality in dielectric nanostructures arises from the constructive and destructive interference of multipole moments with different CPL. By employing generalized multipole decomposition, we introduce a generalized chiral multipole mechanism that allows for direct numerical calculation of CD and establishes the conditions required to achieve maximum chirality. This approach provides a comprehensive framework for analyzing chirality and serves as a foundation for future investigations of chiral nanostructures.

\end{abstract}
\maketitle

Structural chirality refers to the unique property of certain nanostructures, which exhibit asymmetry in their interaction with circularly polarized light due to photonic spin-orbit coupling~\cite{gautier2022planar, zhao2024spin, xiao2017spin, zhang2017symmetry}. This phenomenon arises from the uneven distribution of the electromagnetic field within these structures, significantly influencing their interaction with light~\cite{fernandez2016objects, hendry2012chiral}. The generation of chirality is typically linked to symmetry breaking, both in-plane and out-of-plane, which causes the electromagnetic field to behave asymmetries under different circularly polarized waves~\cite{shi2022planar, wu2021observation}. 
  
Previous research aimed at enhancing chiral responses has largely focused on quantifying the asymmetry in the electromagnetic field and establishing theoretical frameworks to explain mechanisms behind chirality. Key parameters, such as optical chirality density, optical chirality flow, and optical helicity, have been proposed to quantify chirality in nanostructures~\cite{lipkin1964new, kibble1965conservation, przanowski1994some, tang2010optical, cameron2012optical}. Despite these advances, a universal measure of electromagnetic chirality applicable to all fields and structures is still lacking~\cite{tang2010optical}. For instance, while one physical quantity may remain constant for a specific structure, others may exhibit similar variation trends with circular dichroism (CD) as structural parameters change. This inconsistency higlights the mechanisms and conditions that enable a nanostructure to exhibit its strongest chiral response remain underexplored~\cite{chen2024uncovering}. Moreover, most studies focus on quantifying the degree of chirality rather than addressing how to achieve maximum chirality. 

A significant breakthrough in addressing these challenges comes from the recent progress in understanding high-Q optical resonances, particularly quasi-bound states in the continuum (Q-BICs). The discovery and exploration of Q-BICs have provided new insights into how symmetry and high-Q resonances can be leveraged to achieve strong chiral responses in nanostructures  ~\cite{dixon2021self, overvig2021chiral, zhang2022chiral, kuhner2023unlocking, tang2023chiral}. This has, in turn, opened up new directions for research aimed at realizing maximum chirality. However, designing such structures with maximum chirality often requires time-consuming and computationally expensive numerical simulations, emphasizing the need for a generalized physical framework to explain the underlying mechanisms of chirality. Recent theoretical approaches, such as coupled- mode theory (CMT) and reactive helicity density (RHD), have begun to shed light on these mechanisms in specific nanostructures~\cite{chen2024uncovering, gorkunov2020metasurfaces, chen2023observation}.Yet, these methodologies remain limited, as they tend to focus on particular types of chiral structures and fail to establish a general numerical relationship between chirality and the underlying physical parameters.
  
In this letter, for the first time, we reveal that maximum chirality in dielectric nanostructures arises from the constructive and destructive interference between multipole moments interacting with different circularly polarized light. Employing generalized multipole decomposition, we solve the reflection problem via generalized multipole scattering and introduce a generalized chiral multipole mechanism. This mechanim directly correlates chirality with multipoe interactions, enabling us to define generalized  multipole maximum circular dichroism (GMM-CD), thus offering critical insights into maximum chiral responses in dielectric nanostructures.  Numerical simulations on various chiral structures validate our theoretical model, confirming the accuracy and applicability of our proposed framework. 

We consider the chiral nanostructures arrangedperiodically within the x-y plane. Assuming $ e^{-i\omega t} $ time dependence of all fields, the polarization of incident waves can be descirbed  using the complex unit vectors: 
\begin{equation} 
\textbf{e}_\pm = (\textbf{e}_x \mp i\textbf{e}_y)/\sqrt2
\label{equation:1}
\end{equation}
For waves propagating along the negative z direction, $\textbf{e}_+$ and $\textbf{e}_-$ correspond to the right circular polarization (RCP) and left circular polarization (LCP) wave. The denoted + and - represent the results under RCP and LCP wave incident. Additionally, the reflection problem can be explained by Jones matrix. The reflection coefficient can thus be expressed as a two-dimensional column vector:
\begin{equation} 
\textbf{r} = \begin{pmatrix} r_x \\ r_y \end{pmatrix}
\label{equation:2}
\end{equation}
where $r_x$ and $r_y$ represent the amplitude ratio between incident electric field and reflected electric field in x-polarization component and y-polarization component, separately. Subsequently, the reflection coefficient for CPL can be derived from:
\begin{equation} 
\textbf{r}_\pm = \frac{\textbf{e}_x \mp i\textbf{e}_y}{\sqrt2} \cdot \begin{pmatrix} r_x \\ r_y \end{pmatrix} = \frac{1}{\sqrt2}(r_x \mp ir_y)
\label{equation:3}
\end{equation}
The reflection coefficients for right-handed circularly polarized (RCP) and left-handed circularly polarized (LCP) waves, denoted as $\textbf{r}_+$ and $\textbf{r}_-$, respectively, differ primarily in the phase of their y-component, while their magnitudes are identical. This is because circularly polarized light can be viewed as a superposition of two orthogonal linear polarization components (x and y components) with a specific phase relationship. Thus, knowing ($r_x$ and $r_y$) allows calculation of the reflection coefficients for RCP and LCP.   By applying generalized  multipole scattering theory and considering only backward scattering contributions, we derive the reflection coefficients for linearly polarized waves in terms of generalized multipole moments \cite{evlyukhin2016optical, raab2005multipole, landau2013classical, evlyukhin2013multipole, terekhov2019multipole}. For x-polarized incident waves:
\begin{equation}
\begin{split}
  r_x = \frac{ik_d}{2E_{in}S_{p}\varepsilon_0\varepsilon_d}&(\textit{ED}_x - \frac{1}{v_d}\textit{MD}_y + \frac{ik_d}{6}\textit{EQ}_{xz} \\ &- \frac{ik_d}{2v_d}\textit{MQ}_{yz})
\end{split}
\label{equation:4}
\end{equation}
For y-polarized incident waves:
\begin{equation}
\begin{split}
  r_y = \frac{ik_d}{2E_{in}S_{p}\varepsilon_0\varepsilon_d}&(\textit{ED}_y + \frac{1}{v_d}\textit{MD}_x + \frac{ik_d}{6}\textit{EQ}_{yz} \\ &+ \frac{ik_d}{2v_d}\textit{MQ}_{xz})
\end{split}
\label{equation:5}
\end{equation}
where $k_d = k_0\sqrt{\varepsilon_d}$ represents the wave number in a surronding medium ($k_0$ is the wave number in vacuum, $\varepsilon_d$ is the permittivity of sourrouding medium.), $2E_{in}$ is the amplitude of the normally incident plane wave, $\varepsilon_0$ is the vacuum permittivity, and $S_{p}$ is the square lattice. The generalized multipole moments—electric dipole (\textit{ED}), magnetic dipole (\textit{MD}), electric quadrupole (\textit{EQ}), and magnetic quadrupole (\textit{MQ})—are explicitly indicated, with subscripts showing their directional components\cite{ospanova2020generalized, tuz2021polarization, Supplymentary Material}. Subsequently, the reflection coefficient for RCP and LCP waves can be obtained. For RCP waves:
\begin{equation}
\begin{split}
  \textbf{r}_+ = &\frac{ik_d}{2\sqrt{2}E_{in}S_{p}\varepsilon_0\varepsilon_d}((\textit{ED}_x + i\textit{ED}_y) - \frac{1}{v_d}(\textit{MD}_y - i\textit{MD}_x) \\ & + \frac{ik_d}{6}(\textit{EQ}_{xz}-i\textit{EQ}_{yz}) - \frac{ik_d}{2v_d}(\textit{MQ}_{yz} + i\textit{MD}_{xz})
\end{split}
\label{equation:6}
\end{equation}
For LCP waves:
\begin{equation}
\begin{split}
  \textbf{r}_- = &\frac{ik_d}{2\sqrt{2}E_{in}S_{p}\varepsilon_0\varepsilon_d}((\textit{ED}_x - i\textit{ED}_y) - \frac{1}{v_d}(\textit{MD}_y + i\textit{MD}_x) \\ & + \frac{ik_d}{6}(\textit{EQ}_{xz}+i\textit{EQ}_{yz}) - \frac{ik_d}{2v_d}(\textit{MQ}_{yz} - i\textit{MD}_{xz}))
\end{split}
\label{equation:7}
\end{equation}
From Eq. \ref{equation:6} and \ref{equation:7}, The effective polarizability for RCP and LCP can be written as:
\begin{equation}
\begin{split}
\bm{\alpha}^{\mathit{eff}}_\pm &= i \cdot ((\textit{ED}_x \pm i\textit{ED}_y) - \frac{1}{v_d}(\textit{MD}_y \mp i\textit{MD}_x)  + \frac{ik_d}{6}\\ &(\textit{EQ}_{xz} \mp i\textit{EQ}_{yz}) - \frac{ik_d}{2v_d}(\textit{MQ}_{yz} \pm i\textit{MD}_{xz}))E_{in}\varepsilon_0\varepsilon^{\frac{3}{2}}_d
\end{split}
\label{equation:8}
\end{equation}
By interpreting these terms as generalized electric dipole polarizability, generalized magnetic dipole polarizability, generalized electric quadrupole polarizability and generalized magnetic quadrupole polarizability. Then the reflection coefficient can be transformed into:
\begin{equation}
  \mathbf{r}_\pm = \frac{ik_d}{2\sqrt{2}E_{in}S_{p}}(\bm{\alpha}^{\mathit{ED}}_\pm+\bm{\alpha}^{\mathit{MD}}_\pm+\mathbf{\alpha}^{\mathit{EQ}}_\pm+\bm{\alpha}^{\mathit{MQ}}_\pm)
\label{equation:9}
\end{equation}
Then the reflection spectra can be calculated by:
\begin{equation}
\begin{split}
  R_\pm= |\mathbf{r}_\pm|^2 = &\frac{k^2_0}{8E^2_{in}S^2_{p}}|\bm{\alpha}^{\mathit{ED}}_\pm+\bm{\alpha}^{\mathit{MD}}_\pm+\\&\mathbf{\alpha}^{\mathit{EQ}}_\pm+\bm{\alpha}^{\mathit{MQ}}_\pm|^2
\end{split}
\label{equation:10}
\end{equation}

Based on the definition of largest circular dichroism (CD), we can get the CD for reflection spectra\cite{cui2021high, zeng2022giant, tang2022dual}:
\begin{equation}
CD = R_- - R_+
\label{equation:11}
\end{equation}

To achieve the maximum chirality in nanostructure, the reflectance must reach unity for one circular polarization while remaining zero for the opposite polarization. In other words, the dielectric nanostructure should exhibit total reflection for one circular polarization 
\begin{figure}[htbp]
  \centering
  \includegraphics[width=8.6 cm]{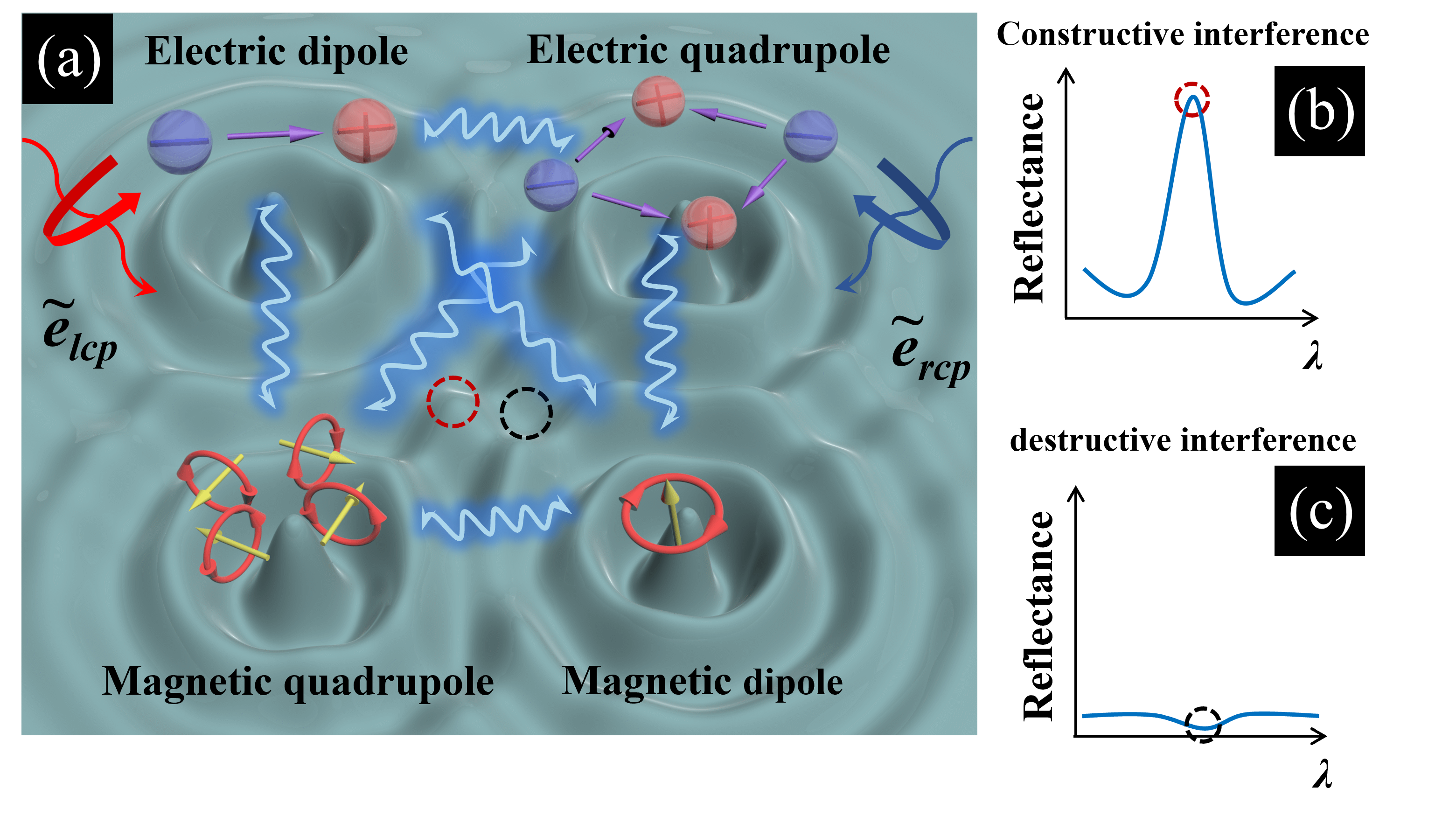}
  \caption{\textbf{The origin of maximum chirality} (a) The concept of generalized chiral multipole mechanism. (b) Constructive interference in reflection spectra. (c) Destructive interference point in reflection spectra.}
\label{fig:1}
\end{figure} 
\begin{figure*}[!t]
  \centering
  \includegraphics[width=17.6cm]{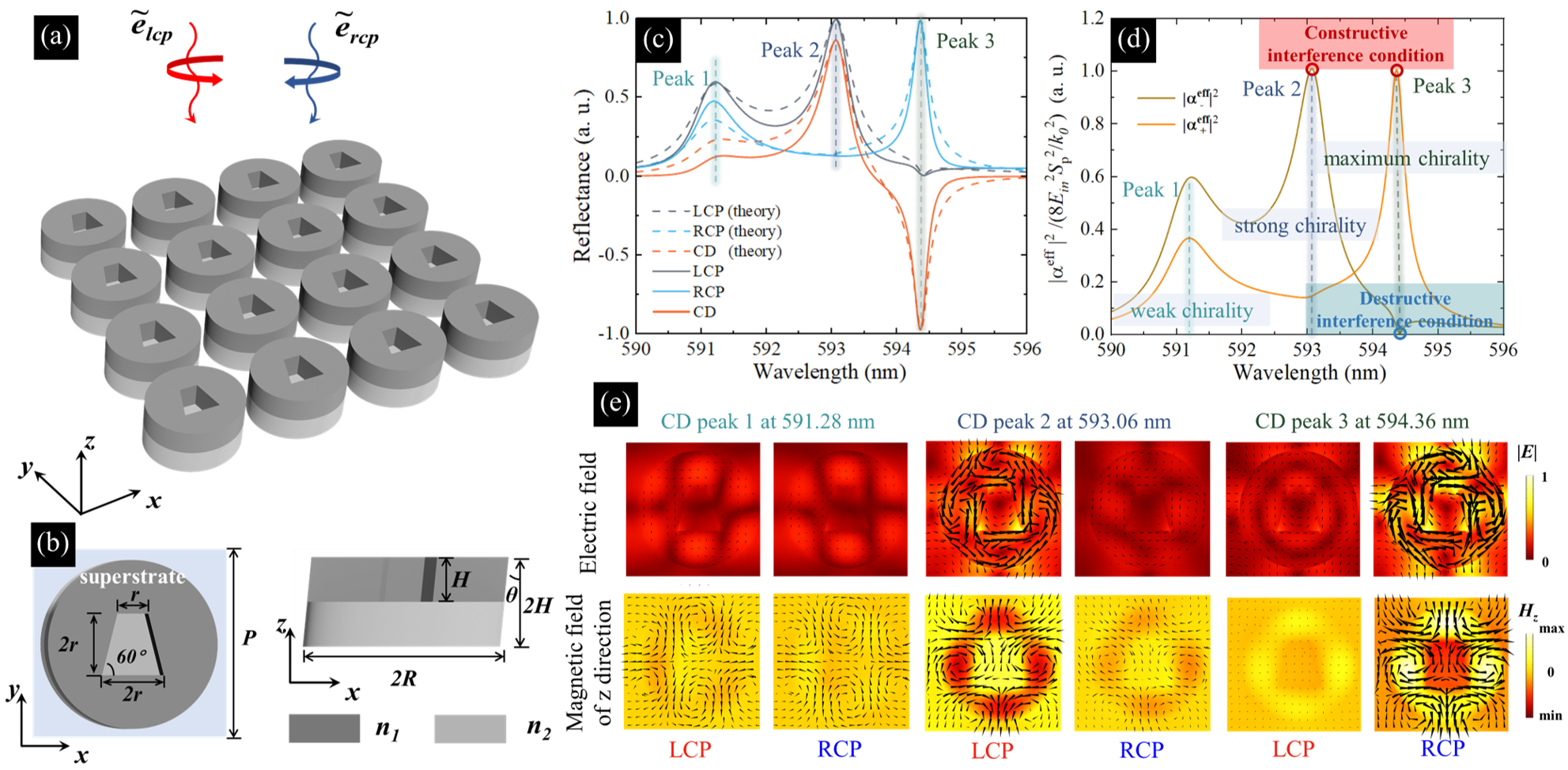}
  \caption{\textbf{Numerical simulation results validating the generalized chiral multipole mechanism}. (a) Illustration of one designed chiral nanostructures. (b) Detail structural parameters of nanostructures. (c) Comparison of reflection and CD spectra from numerical simulation results and theoretical calculation. (d) Effective polarizability for RCP and LCP waves. (e) Electric field (upward) and z-component of the magnetic field (downward) at z = 150 nm for RCP and LCP waves at wavelengths corresponding to weak chirality, strong chirality, and maximum chirality, respectively.}
  \label{fig:2}
  \end{figure*}
and complete transmission (no reflection) for the other at a specific wavelength couldn’t be transmitted for another at one certain wavelength. Finally, we can get the generalized multipole maximum circular dichroism (GMM-CD):

\begin{equation}
\begin{split}
  |\bm{\alpha}^{\mathit{ED}}_\pm+\bm{\alpha}^{\mathit{MD}}_\pm+\mathbf{\alpha}^{\mathit{EQ}}_\pm+\bm{\alpha}^{\mathit{MQ}}_\pm|^2 = \frac{8E^2_{in}S^2_{p}}{k^2_0}\ or \ 0
\end{split}
\label{equation:12}
\end{equation}

The GMM-CD may reveal a simple rule that, as shown by Fig.\ref{fig:1}, if one chiral dielectric nanostructure possess maximum chirality in one certian wavelength, for one circularly polarized wave, the interaction between different multipole moment must forms a constructive interference, causing the total reflectance. For another, the interaction must forms a destructive interference, leading to the state of total transmission, also known as anapole effect \cite{savinov2019optical, gurvitz2019high, colom2019modal, allayarov2024anapole, yang2019nonradiating}.

To validate our proposed generalized chiral multipole mechanism, we numerically  conducted numerical simulations of various chiral nanostructures using COMSOL MULTIPHYSICS (additional simulations are provided in the supplymentary Material.(\textbf{IV}))\cite{Supplymentary Material}. Here, we discuss one specific chiral structure depicted in Fig. \ref{fig:2} (a) and (b). The unit cell of this nanostructure consists of a tilted double-layer cylinder with a tilted trapezoidal hole in the superstrate. The period P is 500 nm, the tilt angle $\theta$ is 8 degrees, the radius R of the cylinder is 220 nm, and both the thickness H of each layer and the depth of the trapezoidal hole are 100 nm. The refractive index $n_1$ of the superstrate is 3.44, and the refractive index $n_2$ of the substrate is 1.59. For simplicity, the permittivity of the surrounding medium is set to 1.

First, we simulated the reflection for normally incident RCP and LCP waves and calculate the corresponding CD spectra of the reflected light. As shown in Fig. \ref{fig:2}(c), the chiral nanostructure exhibits three distinct resonant modes within the wavelength range of 590 nm to 596 nm, under both RCP and LCP wave illumination. TThese resonant behaviors contribute significantly to chirality. According to Eq. \ref{equation:11}, the chiral nanostructure shows three chiral resonant states within this wavelength range. At 591.28 nm, the first CD peak emerges, showing slightly higher reflection for LCP compared to RCP, resulting in a CD value around 0.16. The second peak at 593.06 nm displays total reflection for LCP waves and dominant transmission for RCP waves, leading to a CD value of approximately 0.83. The third peak at 594.36 nm shows total reflection for RCP waves, highlighting strong chirality and achieving maximum CD. Additionally, the far-field polarization maps provided in \ref{fig:S2} illustrate polarization state distributions in k-space for these resonances, and the calculated AN-S3 /S0  correspond closely with the simulation results (Details see in Supplymentary Material IV) \cite{Supplymentary Material}.

\begin{figure*}[!t]
  \centering
  \includegraphics[width=17.6cm]{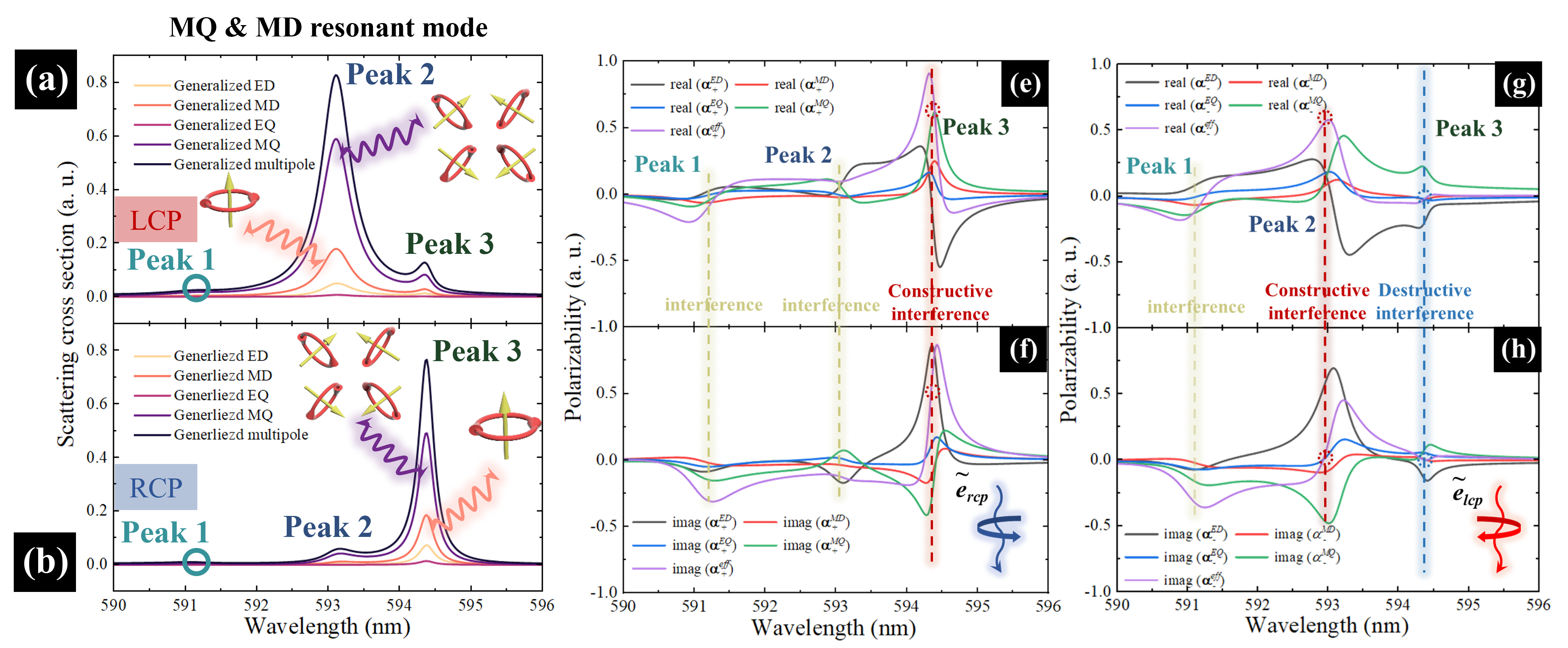}
  \caption{\textbf{Generalized multipole distribution for multipole scattering and effective polarizability for each generalized multipole under RCP and LCP waves.}. (a) Generalized multipole scattering cross section under LCP waves. (b) Generalized multipole scattering under RCP waves. (c) The real part of each Generalized multipole effective polarizability under RCP waves. (d) The imaginary part of each Generalized multipole effective polarizability under RCP waves. (e) The real part of each Generalized multipole effective polarizability under LCP waves. (f) The imaginary part of each Generalized multipole effective polarizability under LCP waves.}
  \label{fig:3}
\end{figure*}
Subsequently, we apply our generalized chiral multipole mechanism to calculate the reflection and CD spectra based on Eq. \ref{equation:6} and Eq. \ref{equation:7}. As depicted in Fig. \ref{fig:2}(c), the theoretically obtained spectra agree closely with the numerical simulation results, thereby confirming the effectiveness and reliability of our proposed framework.

To further validate the principles revealed by the GMM-CD, we calculated the effective polarizability for RCP and LCP waves using Eq. \ref{equation:12}, and assessed whether they follow the GMM-CD rules across the three chiral resonant states. As shown in Fig. \ref{fig:2}(d), at CD peak 1, the effective polarizability for both RCP and LCP waves does neither satisfy the constructive or destructive interference conditions, but still shows a difference, resulting in weak chirality. At CD peak 2, the effective polarizability for LCP waves demonstrates constructive interference condition, while the effective polarizability for RCP waves reflects a weak destructive interference state between 
the generalized multipole moments, leading to strong chirality. At CD peak 3, the effective polarizability for RCP waves satisfies the constructive interference condition, while the effective polarizability for LCP waves satisfies the destructive interference condition, corresponding to the maximum chirality point. The electric field (up) and z-component of magnetic field (down) for three chiral resonant states is depicted in Fig. \ref{fig:2} (e). The arrows for upward illustrate represent the planar current distribution while for downward represent the planar magnetic field distribution. The differences between the electromagnetic responses under LCP and RCP illumination become increasingly pronounced from the first to the third peak. Notably, at the maximum chirality point (CD peak at 594.36 nm), the electromagnetic response under LCP illumination is significantly weaker, corresponding clearly to destructive interference between multipole moments.

To show case the superiority of our model, we compared our generalized chiral multipole mechanism with the traditional generalized multipole scattering method, which extensively applied in many works to investigate the resonant states and study how they induce the chirality with our generalized chiral multipole mechanism \cite{yang2019ultrahighly, cai2021symmetric, zhong2024toroidal, zhu2024merging, yangultrasensitive}.   
The results of generalized multipole scattering are depicted in Fig. \ref{fig:3} (a)-(b) and effective polarizability based on generalized chiral multipole mechanism are depicted in Fig. \ref{fig:3} (c)-(f). 
Results from the conventional multipole scattering analysis, presented in Fig. \ref{fig:3} (a) and (b), , indicate that resonant states under RCP and LCP illumination are predominantly governed by \textit{MQ} and \textit{MD} modes, with minimal contributions from other multipoles. Differences mainly arise in intensity at constructive and destructive interference points(details see in Supplymentary Material)\cite{Supplymentary Material}. And there is only difference in intensity when it is comes to the destructive interference point compared with constructive interference point.
Intriguingly, for effectively polarizability of different multipole moments under RCP and LCP waves, the constructive interference is dominated by the \textit{MQ} and \textit{ED} mode, which differs from the results of generalized multipole scattering[see constructive interference points in Fig. \ref{fig:3} (c)-(f)]. At destructive interference point, the intensity of effective polarizability for \textit{MQ} and \textit{ED} mode  is still high but their signs are opposite while other multipole moments remian near zero [see destructive interference points in Fig. \ref{fig:3} (e) and (f)]. The different results imply that the focuses of two physic model are different.
The generalized multipole scattering primarily captures the global scattering behavior ,incorporating all multipole components(details see in Supplymentary Material.(\textbf{II})\cite{Supplymentary Material}). Relying on Eq. (\ref{equation:6}) and Eq. (\ref{equation:7}), the resonance peaks in reflection spectrum actually mostly depend on some particular components of generalized multipole moments. Therefore, it cannot precisely identify specific interactions causing resonances or accurately explain the generation of chirality. Our generalized chiral multipole approach, however, emphasizes specific multipole components interactions that determine resonance peaks and effectively clarifies the mechanisms driving strong chirality.

In summary, we uncover the physical origin of maximum chirality through multipole interference and propose the generalized chiral multipole mechanism (GMM-CD). Constructive interference induces total reflectance for one circular polarization, while destructive interference (anapole effect) leads to total transmission for the opposite polarization. By computing the CD spectrum from specific multipole components—validated against simulations—we identify the dominant contributors to chirality, surpassing conventional multipole scattering analysis. Our framework demonstrates robust universality and establishes a direct numerical relation between multipole interference and chirality. These findings offer a comprehensive understanding of chirality mechanisms, setting a solid foundation for further research in chiral dielectric nanostructures.

\section*{\textbf{Acknowledge}}
This work is supported by National Natural Science Foundation of China (62104174, 62205253) and Hubei Provincial Natural Science Foundation of China (2021CFB054).

\textsuperscript{†} These authors contributed equally to this work.

\textsuperscript{†}:\hypertarget{note1}{340625@whut.edu.cn}

\textsuperscript{†}:\hypertarget{note2}{shengyiwang@whut.edu.cn}

\renewcommand{\theequation}{S\arabic{equation}}
\renewcommand{\thefigure}{S\arabic{figure}}
\setcounter{equation}{0} 
\setcounter{figure}{0}   
\onecolumngrid
\newpage
\begin{center}
    {\large \textbf{Supplementary Material for: \\[0.5em] Origin of Maximum chirality optical resonance}}\\[1.5em]
    {\large WenKui Zhao\textsuperscript{1,{\hyperlink{note1}{†}}}, ShengYi Wang\textsuperscript{1,{\hyperlink{note2}{†}}}, Hanzhuo Kuang\textsuperscript{1}, Luo Hao\textsuperscript{1}, Qiu Wang\textsuperscript{1} and Bo-Wen Jia\textsuperscript{1,*}}\\[1em]
    \textit{
    \textsuperscript{1}School of Information Engineering, Wuhan University of Technology, Wuhan 430070, China\\
    \textsuperscript{2}Key Laboratory of Artificial Micro- and Nano-structures of Ministry of Education, and School of Physical and Technology, Wuhan University, Wuhan 430072, China}\\

\end{center}
\vspace{-0.5em} 
The Supplymentary Material includes the following sections:(\textbf{I.}) Detail derivation for generalized chiral multipole mechansim. (\textbf{II.}) Generalized multipole scattering cross section. (\textbf{III.}) Electromagnetic field for the chiral nanostructures in text. (\textbf{IV.}) Far field polarization maps for three peaks. (\textbf{V.}) Bilayer cylinder metasurface. (\textbf{VI.}) Bilayer photonic crystal.
\vspace{1em}
\section*{\textbf{I.}Detail derivation for generalized chiral multipole mechanism.}
To analyze chiral nanostructures in the x-y plane arranged periodically, we consider waves with time dependence \( e^{-i\omega t} \), polarized following complex unit vectors:
\begin{equation} 
\textbf{e}_\pm = (\textbf{e}_x \mp i\textbf{e}_y)/\sqrt{2}
\label{equation:S1}
\end{equation}
For waves propagating along the negative z-direction, \( \textbf{e}_+ \) and \( \textbf{e}_- \) correspond to right (RCP) and left (LCP) circular polarization. The reflection coefficients for RCP and LCP waves are derived from the Jones matrix representation of the reflection coefficients \( r_x \) and \( r_y \), defined as:
\begin{equation} 
\textbf{r} = \begin{pmatrix} r_x \\ r_y \end{pmatrix}
\label{equation:S2}
\end{equation}
These represent the amplitude ratios of incident and reflected electric fields in x- and y-polarizations. Using these coefficients, the reflection coefficients for CPL are given by:
\begin{equation} 
\textbf{r}_\pm = \frac{\textbf{e}_x \mp i\textbf{e}_y}{\sqrt{2}} \cdot \begin{pmatrix} r_x \\ r_y \end{pmatrix} = \frac{1}{\sqrt{2}}(r_x \mp ir_y)
\label{equation:S3}
\end{equation}
These coefficients reflect the phase relationship inherent in circularly polarized light as a combination of orthogonal linear polarizations. The reflection and transmission problem are strongly related to the electromagnetic scattering performance, which can be described by multipole scattering \cite{evlyukhin2016optical, raab2005multipole, landau2013classical, evlyukhin2013multipole, terekhov2019multipole}. If the dipole moments, quadrupole moments, mean-square radii are all taken into account, the generalized multipole moments can be derived \cite{ospanova2020generalized, tuz2021polarization}. The dipole moments, quadrupole moments and homologous mean-square radii interference and can induce hybrid anapole effect, the generalized multipole moments are proposed to  capture this effect and provide a more precise basis to solve the reflection and transmission problem.

\begin{table}[h!]
  \centering
  \renewcommand{\arraystretch}{1.5} 
  \begin{tabular}{p{0.25\textwidth} p{0.7\textwidth}}
  \toprule
  \textbf{Category} & \textbf{Expression} \\ 
  \midrule
  Dipole moments & 
  $p_i = \frac{i}{\omega} \int j \, \mathrm{d}^3 r, \quad
  m_i = \frac{1}{2} \int (\textbf{r} \times \textbf{j}) \, \mathrm{d}^3 r, \quad
  T_i = \frac{1}{10} \int [(r \cdot j)r_i - 2r^2 j_i] \, \mathrm{d}^3 r$ \\ 
  
  Quadrupole moments & 
  \(
  \begin{array}{l}
  Q_{ij}^e = \frac{i}{\omega} \int_V [r_i j_j + r_j j_i - \frac{2}{3} \delta_{ij} (\textbf{r} \cdot \textbf{j}) r_i] \, \mathrm{d}^3 r, \quad
  Q_{ij}^m = \frac{1}{3} \int_V [(\textbf{r} \times \textbf{j})_i r_j + (\textbf{r} \times \textbf{j})_j r_i] \, \mathrm{d}^3 r, \\
  Q_{ij}^T = \frac{1}{42} \int_V [4(\textbf{r} \cdot \textbf{j})r_i r_j + 2(\textbf{j} \cdot \textbf{r})r^2 \delta_{ij} - 5r^2 (r_i j_j + r_j j_i)] \, \mathrm{d}^3 r
  \end{array}
  \) \\

  Mean-square radii & 
  \(
  \begin{array}{l}
  M_i^{\langle 2 \rangle} = \frac{i \omega}{20} \int r^2 (\textbf{r} \times \textbf{j}) \, \mathrm{d}^3 r, \quad
  T_i^{\langle 2 \rangle} = \frac{1}{280} \int [3r^4 j_i - 2r^2 (r \cdot j)r_i] \, \mathrm{d}^3 r, \\
  Q_{ij}^{m\langle 2 \rangle} = \frac{i \omega}{42} \int r^2 [(\textbf{r} \times \textbf{j})_i r_j + (\textbf{r} \times \textbf{j})_j r_i] \, \mathrm{d}^3 r
  \end{array}
  \) \\  
  
  Generalized multipole &
  \(
  \begin{array}{l}
  ED_i = p_i + \frac{ik^2 \varepsilon_d}{c} T_i + \frac{ik^3 \varepsilon_d^2}{c} T_i^{\langle 2 \rangle}, \quad
  MD_i = m_i + \frac{ik \varepsilon_d}{c} M_i^{\langle 2 \rangle}, \\
  EQ_i = Q_{ij}^e + \frac{ik \varepsilon_d}{c} Q_{ij}^T, \quad
  MQ_i = Q_{ij}^m + \frac{ik \varepsilon_d}{c} Q_{ij}^{(2)}
  \end{array}
  \) \\ 
  \bottomrule
  \end{tabular}
  \caption{Mathematical expressions for dipole moments, quadrupole moments, mean-square radii, and generalized multipole.}
  \label{table:expressions}
\end{table}

The mathematical expressions are depicted in table. \ref{table:expressions}, where $p_i$, $m_i$ and $T_i$ represent the component of electric dipole, magnetic dipole and toroidal dipole, $Q_{ij}^e$ and $Q_{ij}^m$ and $Q_{ij}^T$ represent the component of electric quadrupole, magnetic quadrupole and toroidal quadrupole moment, $ M_i^{\langle 2 \rangle} $, $ T_i^{\langle 2 \rangle} $ and $ Q_{ij}^{m\langle 2 \rangle} $ represent the component of mean-square radii for magnetic dipole, toroidal dipole and magnetic quadrupole moment, the $ED_i$, $MD_i$, $EQ_i$ and $MQ_i$ represent the component of generalized electric dipole, magnetic dipole, electric quadrupole and magnetic quadrupole moments. $ k $ is the wave vector, $ \varepsilon_d $ is the permittivity of a surrounding medium and $ c $ is the speed of light in vacuum. Note that these expressions are approximations when the structural periodic size is smaller than the wavelength, if getting the more precise results, the n-order spherical Bessel function must be taken into account.

The scattering field can be approximated by generalized multipole moments \cite{evlyukhin2016optical, raab2005multipole, landau2013classical, evlyukhin2013multipole, terekhov2019multipole}: 

\begin{equation}
\mathbf{E}^{\text{sc}}(\mathit{n}) \sim 
\left(
    [\mathit{n} \times (\mathit{ED} \times \mathbf{n})] 
    + \frac{1}{v_d} [\mathit{MD} \times \mathbf{n}]
    + \frac{i k_d}{6} [\mathbf{n} \times (\mathbf{n} \times \hat{\mathit{EQ}} \mathbf{n})]
    + \frac{i k_d}{2 v_d} [\mathbf{n} \times (\hat{\mathit{MQ}} \mathbf{n})] \}
\right),
\label{equation:S4}
\end{equation}
where $\mathbf{n} $ is the unit vector indicating the scattering direction, $\mathit{ED}$ and $\mathit{MD}$ are generalized electric dipole and magnetic dipole moment, $\hat{\mathit{EQ}}$ and $\hat{\mathit{MQ}}$ are the symmetrized and traceless tensors of generalized electric quadrupole and magnetic quadrupole moments. $k_d$ is the wave vector in the surrouding medium and $v_d$ is the speed of light in the surrouding medium. If we approximate the refelction and transmission problem by backward and forward scattering of generalized multipole moments that we inserting $\mathbf{n} = (0,0,n_z)$ where $n_z = 1$ or $-1$ into the Equation.\ref{equation:S4} to calculate the reflection and transmission coefficient. Thus, in the case of x-polarization:

\begin{align}
  r_x &= \frac{ik_d}{E_{in} 2 S_p \varepsilon_0 \varepsilon_d} 
      \left( 
          ED_x - \frac{1}{v_d} MD_y + \frac{ik_d}{6} EQ_{xz} 
          - \frac{ik_d}{2v_d} MQ_{yz} 
      \right), \label{equation:S5} \\
  t_x &= 1 + \frac{ik_d}{E_{in} 2 S_p \varepsilon_0 \varepsilon_d} 
      \left( 
          ED_x + \frac{1}{v_d} MD_y - \frac{ik_d}{6} EQ_{xz} 
          - \frac{ik_d}{2v_d} MQ_{yz} 
      \right), \label{equation:S6}
  \end{align}
  
in the case of y-polarization:

\begin{align}
  r_y &= \frac{ik_d}{E_{in} 2 S_p \varepsilon_0 \varepsilon_d} 
      \left( 
          ED_y + \frac{1}{v_d} MD_x + \frac{ik_d}{6} EQ_{yz} 
          + \frac{ik_d}{2v_d} MQ_{xz} 
      \right), \label{equation:S7}\\
  t_y &= 1 + \frac{ik_d}{E_{in} 2 S_p \varepsilon_0 \varepsilon_d} 
      \left( 
          ED_y - \frac{1}{v_d} MD_x - \frac{ik_d}{6} EQ_{yz} 
          + \frac{ik_d}{2v_d} MQ_{xz} 
      \right). \label{equation:S8}
  \end{align}
  
  Subsequently, we only consider the reflection problem due to for dielectric nanostructures, the reflection problem is actually equal to transmission problem. Inserting the Eq. \ref{equation:S5} and \ref{equation:S7} into \ref{equation:S3}, the reflection coefficient for circularly polarized waves can be obtained, for RCP waves:
  \begin{equation}
  \begin{split}
    \textbf{r}_+ = \frac{ik_d}{2\sqrt{2}E_{in}S_{p}\varepsilon_0\varepsilon_d}((\textit{ED}_x - i\textit{ED}_y) - \frac{1}{v_d}(\textit{MD}_y + i\textit{MD}_x)  + \frac{ik_d}{6}(\textit{EQ}_{xz}+i\textit{EQ}_{yz}) - \frac{ik_d}{2v_d}(\textit{MQ}_{yz} - i\textit{MD}_{xz})),
  \end{split}
  \label{equation:S9}
  \end{equation}
for LCP waves:
  \begin{equation}
  \begin{split}
    \textbf{r}_- = \frac{ik_d}{2\sqrt{2}E_{in}S_{p}\varepsilon_0\varepsilon_d}((\textit{ED}_x + i\textit{ED}_y) - \frac{1}{v_d}(\textit{MD}_y - i\textit{MD}_x)  + \frac{ik_d}{6}(\textit{EQ}_{xz}-i\textit{EQ}_{yz}) - \frac{ik_d}{2v_d}(\textit{MQ}_{yz} + i\textit{MD}_{xz}))
  \end{split}
  \label{equation:S10}
  \end{equation}
  From Eqs.\ref{equation:S9} and \ref{equation:S10}, the effective polarizability for RCP and LCP can be expressed as:
  \begin{equation}
  \begin{split}
  \bm{\alpha}^{\mathit{eff}}_\pm = i \cdot ((\textit{ED}_x \mp i\textit{ED}_y) - \frac{1}{v_d}(\textit{MD}_y \pm i\textit{MD}_x)  + \frac{ik_d}{6} 
  (\textit{EQ}_{xz} \pm i\textit{EQ}_{yz}) - \frac{ik_d}{2v_d}(\textit{MQ}_{yz} \mp i\textit{MD}_{xz}))E_{in}\varepsilon_0\varepsilon^{\frac{3}{2}}_d
  \end{split}
  \label{equation:S11}
  \end{equation}

  Here, the terms correspond to generalized electric dipole, magnetic dipole, electric quadrupole, and magnetic quadrupole polarizabilities (see in table. \ref{table:polarizability}). 
\begin{table}[h]
    \centering
    \renewcommand{\arraystretch}{1.5}
    \begin{tabular}{ll}
        \toprule
        \textbf{Category} & \textbf{Expression} \\ 
        \midrule
        Generalized electric dipole polarizability ($\bm{\alpha}^{\mathit{ED}}_\pm$) & 
        $ \bm{\alpha}^{\mathit{ED}}_\pm = i(\textit{ED}_x \mp i\textit{ED}_y)$ \\ 
        
        Generalized magnetic dipole polarizability ($\bm{\alpha}^{\mathit{MD}}_\pm$) & 
        $ \bm{\alpha}^{\mathit{MD}}_\pm = -\frac{i}{v_d}(\textit{MD}_y \pm i\textit{MD}_x)$ \\ 
        
        Generalized electric quadrupole polarizability ($\bm{\alpha}^{\mathit{EQ}}_\pm$) & 
        $ \bm{\alpha}^{\mathit{EQ}}_\pm = \frac{ik_d}{6}(\textit{EQ}_{xz} \pm i\textit{EQ}_{yz})$ \\ 
        
        Generalized magnetic quadrupole polarizability ($\bm{\alpha}^{\mathit{MQ}}_\pm$) & 
        $ \bm{\alpha}^{\mathit{MQ}}_\pm = -\frac{ik_d}{2v_d}(\textit{MQ}_{yz} \mp i\textit{MQ}_{xz})$ \\ 
        \bottomrule
    \end{tabular}
    \caption{Effective polarizability components for RCP and LCP waves.}
    \label{table:polarizability}
\end{table}

Then, the reflection coefficient and reflection spectra becomes:
  \begin{equation}
    \mathbf{r}_\pm = \frac{ik_d}{2\sqrt{2}E_{in}S_{p}}(\bm{\alpha}^{\mathit{ED}}_\pm+\bm{\alpha}^{\mathit{MD}}_\pm+\mathbf{\alpha}^{\mathit{EQ}}_\pm+\bm{\alpha}^{\mathit{MQ}}_\pm)
  \label{equation:S12}
  \end{equation}

  \begin{equation}
  \begin{split}
    R_\pm= |\mathbf{r}_\pm|^2 = \frac{k^2_0}{8E^2_{in}S^2_{p}}|\bm{\alpha}^{\mathit{ED}}_\pm+\bm{\alpha}^{\mathit{MD}}_\pm+
    \mathbf{\alpha}^{\mathit{EQ}}_\pm+\bm{\alpha}^{\mathit{MQ}}_\pm|^2
  \end{split}
  \label{equation:S13}
  \end{equation}
  The largest circular dichroism (CD) for the reflection spectra is defined as:
  \begin{equation}
  CD = R_- - R_+
  \label{equation:S14}
  \end{equation}

  To achieve maximum chirality in a nanostructure, the reflectance under one circularly polarized wave must reach unity while the other remains zero. This indicates that the dielectric nanostructure achieves total transmission for one circularly polarized wave while completely blocking the other at a specific wavelength. The generalized multipole maximum circular dichroism (GMM-CD) can be expressed as:
\begin{equation}
\begin{split}
  |\bm{\alpha}^{\mathit{ED}}_\pm+\bm{\alpha}^{\mathit{MD}}_\pm+\mathbf{\alpha}^{\mathit{EQ}}_\pm+\bm{\alpha}^{\mathit{MQ}}_\pm|^2 = \frac{8E^2_{in}S^2_{p}}{k^2_0} \quad \text{or} \quad 0,
\end{split}
\label{equation:S15}
\end{equation}
as shown by Eq.\ref{equation:S15}, the GMM-CD may reveal a simple rule: for a chiral dielectric nanostructure exhibiting maximum chirality at a specific wavelength, the interaction between different multipole moments forms constructive interference for one circularly polarized wave, resulting in total reflectance. Conversely, for the other circularly polarized wave, the interaction forms destructive interference, leading to total transmission—a phenomenon also known as the anapole effect.

\section*{\textbf{II.}Generalized multipole scattering cross section}
Considering the first few spherical multipoles in the Cartesian basis, the scattering cross section (SCS), defined as the ratio of scattered power to the incident wave intensity, can be expressed as a sum of contributions from electric dipoles, magnetic dipoles, and higher-order multipoles. Considering the generalized multipole moments, the scattering cross section can be obtained:
\begin{equation}
  \sigma_{\text{sca}} \simeq 
  \frac{k_0^4}{6\pi\varepsilon_0^2 |\mathbf{E}|^2} |\mathit{ED}|^2 
  + \frac{k_0^6 \varepsilon_s}{720 \pi \varepsilon_0^2 |\mathbf{E}|^2} \sum_{\alpha\beta} |\mathit{EQ}_{\alpha\beta}|^2 
  + \frac{k_0^4 \varepsilon_d \mu_0}{6\pi\varepsilon_0 |\mathbf{E}|^2} |\mathit{MD}|^2 
  + \frac{k_0^6 \varepsilon_d^2 \mu_0}{80 \pi \epsilon_0 |\mathbf{E}|^2} \sum_{\alpha\beta} |\mathit{MQ}_{\alpha\beta}|^2, 
  \label{equation:S16}
  \end{equation}
These multipole moments can be calculated by the mathematical expressions in table. \ref{table:expressions}. The generalized multipole scattering cross section theory is a universal approach that primarily focuses on the overall scattering behavior of different multipole moments. The scattering cross section declines when dipole moments, quadrupole moments and mean-square radii interferences forming a hybrid anapole states. In this framework, all components of various multipole moments—such as dipole moment, quadrupole moments and mean-square radii are included in the calculations, providing a comprehensive description of the scattering properties of nanostructures. However, the main limitation of this approach lies in its inability to precisely explain the underlying mechanisms of resonance peaks in the reflection spectrum or to reflect the interactions between different multipole moments. Specifically, the physical origin of resonance peaks often depends on specific components of the multipole moments rather than their aggregate contributions. 

Based on the analysis of Eq. (\ref{equation:S5}) and Eq. (\ref{equation:S7}), it becomes evident that the resonance peaks in the reflection spectrum are primarily determined by certain particular components of the generalized multipole moments. This indicates that, while the generalized multipole scattering theory effectively describes the global scattering behavior, it is less suitable for investigating how individual multipole moments induce strong resonances and contribute to the generation of chirality. Thus, this theory falls short in uncovering the fundamental mechanisms underlying the enhancement of chirality.

The generalized chiral multipole mechanism we proposed offers a more targeted and applicable method. Compared to traditional approaches, this mechanism accurately captures the interactions between specific components of the multipole moments and elucidates how these interactions lead to resonance enhancement and the emergence of chirality. This new mechanism not only bridges the gap left by the generalized multipole scattering theory but also provides a novel theoretical tool for exploring the role of multipole moments in nanophotonics and chiral optics. By introducing this approach, we pave the way for a deeper understanding of the physical mechanisms behind chirality and for the design of nanostructures with tailored optical properties.

\section*{(\textbf{III.}) Resonant mode for the chiral nanostructures in text.}

\begin{figure*}[h]
  \centering
  \includegraphics[width=17.6cm]{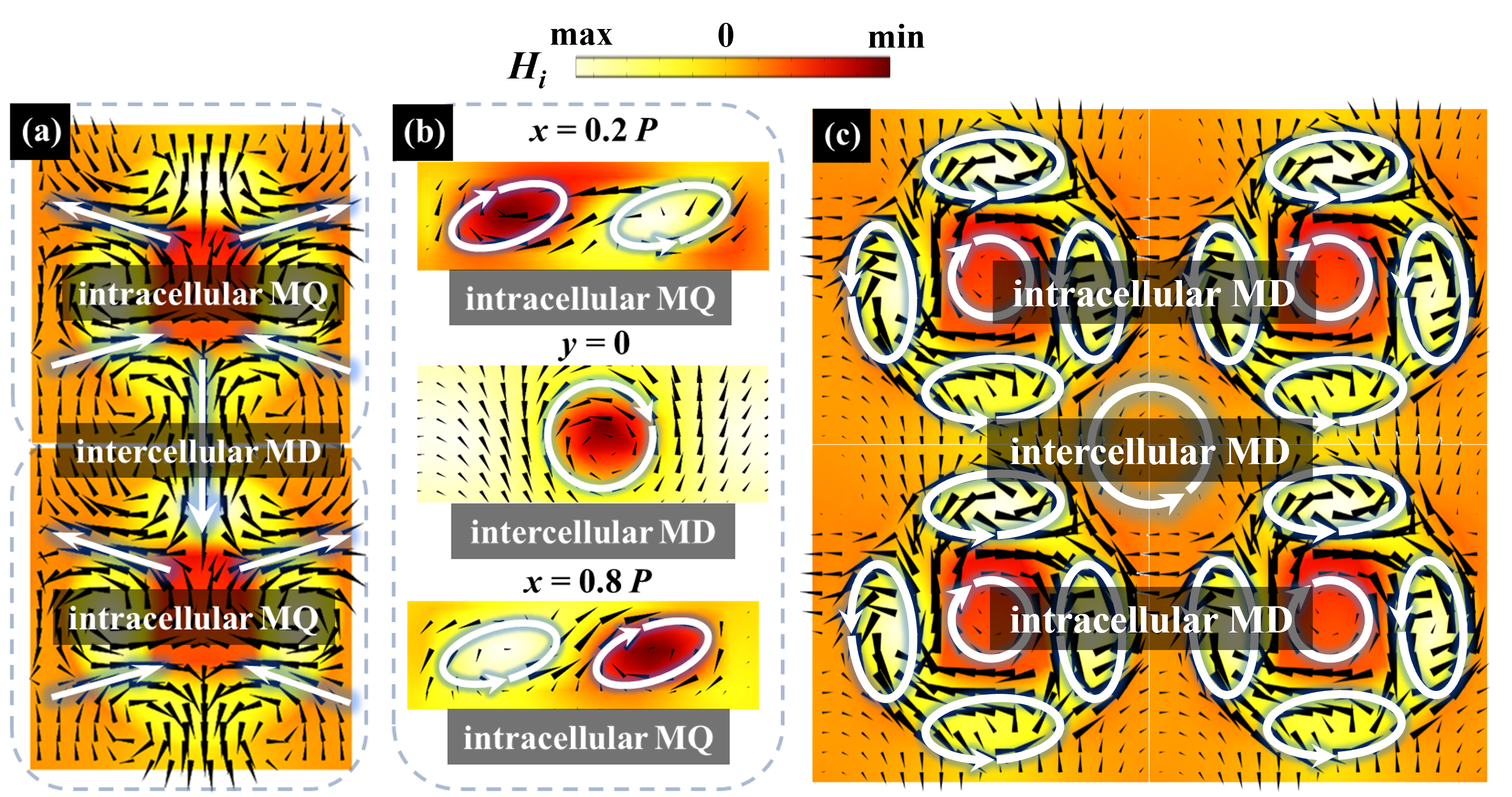}
  \caption{\textbf{Resonant mode for chiral nanostructures proposed in text}. (a) z-component of magnetic field in z = H plane where array present the planar distribution for magnetic field (b) Vertical component of magnetic field for different plane where array present the planar component of displacement current. (c) z-component for magnetic field in z = H plane where array present the planar component of displace current. }
  \label{fig:S1}
\end{figure*}

As shown in Fig.\ref{fig:3} (a), the proposed chiral nanostrucrtures are dominated by generalized MQ and MD mode. Here in Fig. \ref{fig:S1}, we represent vertical component of magnetic field and planar component of magnetic field ( see in Fig. \ref{fig:S1} (a) )displacement current ( see in Fig. \ref{fig:S1} (b) and (c) ) for different planes. In Fig. \ref{fig:S1} (a), it is shown that the chiral nanostructures form intracellular MQ and intercellular MD mode. To further validate the existence of intracellur MQ and intercellular MD mode, vertical component of magnetic field for $x = 0.2P$ , $ y = 0$ and $ x = 0.8P $ planes where array present the planar component of displace current are depicted in Fig. \ref{fig:S1} (b). In $x = 0.2P$ and $x = 0.8P$ plane, the displacement current vibrates and forms two current loop following different direction of rotation, which correspond to two unparallel magnetic dipoles. The total four MD form the intracellur MQ. In y = 0 plane, it is clear to see that the displacement current oscillates and form one current loop corresponding to one intercellular MD mode following negative y-direction, confirming the existence of intercellular MD mode. In this section, we only present the field distribution at 594.36 nm under right circularly polarized (RCP) waves. The field distributions for other resonant peaks are similar; therefore, they are not displayed separately.
\section*{(\textbf{IV.}) Far field polarization maps for three peaks.}
\begin{figure*}[htbp]
  \centering
  \includegraphics[width=17.6cm]{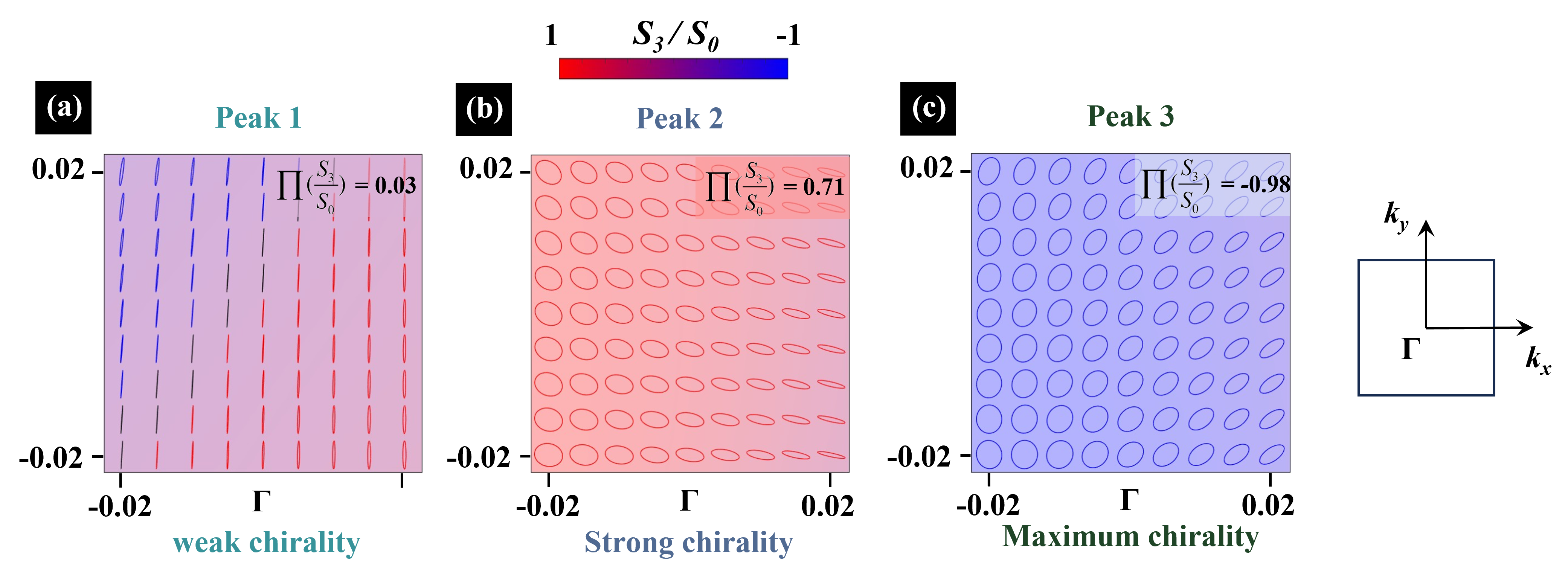}
  \caption{\textbf{Far field polarization maps for three peaks}. (a) Far field polarization map for weak chirality peak. (b) Far field polarization map for strong chirality peak. (c) Far field polarization map for maximum chirality peak. The red and blue ellipses correspond to left and right polarization states.}
  \label{fig:S2}
\end{figure*}
Fig. \ref{fig:S2} illustrates the far-field polarization maps for three distinct peaks. The background color represents the variation in polarization states, while the ellipticity of the ellipses indicates the degree of polarization ranging from linear to circular. Using Eq. \ref{equation:S17}, the average normalized $S_3/S_0$ parameter (AN-$S_3/S_0$) is computed and displayed in the respective polarization maps.
\begin{equation}
  \prod \left( \frac{S_3}{S_0} \right) = \frac{1}{S_k} \iint_{k\text{-space}} \frac{S_3}{S_0} \, dk
  \label{equation:S17}
  \end{equation}
For Peak 1, which exhibits weak chirality, the ellipticity in the k-space predominantly maintains a linear polarization state, resulting in an AN-$S_3/S_0$ parameter close to 0. In contrast, Peak 2, characterized by strong chirality, demonstrates right-handed polarization states, with the calculated AN-$S_3/S_0$ parameter reaching 0.71, consistent with the strong chirality condition. Finally, for the maximum chirality peak, the ellipticity in the k-space almost entirely adopts a right-handed polarization state, and the AN-$S_3/S_0$ parameter approaches nearly -1. The far field polarization maps fit well with the simulation results and reflect the chirality dependence with the direction of wavevector, paving a way for adjusting the chirality by introducing the intrinsic chirality together with the extrinsic chirality.

\section*{(\textbf{V.}) Bilayer cylinder metasurface.}

Here we show that one nanostructue even without structural chirality, the generalized chiral multipole mechanism is still applicable to calculate the reflection and transmission spectra. 

\begin{figure*}[htbp]
  \centering
  \includegraphics[width=13.6cm]{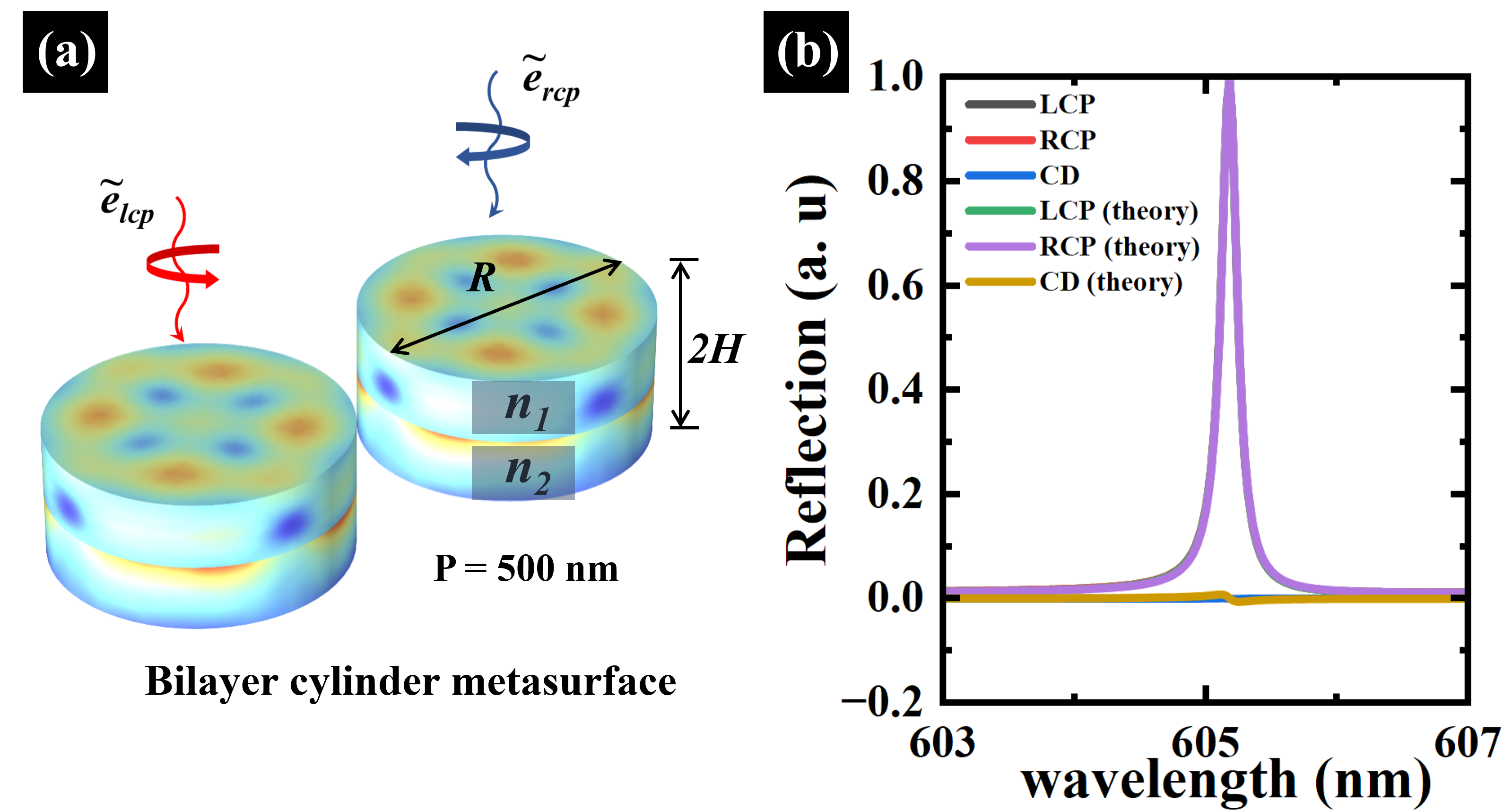}
  \caption{\textbf{Bilayer cylinder metasurface}. The sketch of near fields (interms of total electric field amplitude $|\textbf{E}|$)at resonant point of bilayer cylinder under RCP and LCP waves. }
  \label{fig:S3}
\end{figure*}

In this section, we investigate the generalized chiral multipole mechanism in a bilayer cylinder metasurface. As illustrated in Fig. \ref{fig:S3}(a), the unit cell of the bilayer cylinder metasurface consists of two cylindrical layers, with the superstrate having a refractive index of 3.44 and the substrate a refractive index of 1.59. The period P is 500 nm, the radius R is 220 nm, and the thickness H is 100 nm. The near fields of the bilayer cylinder exhibit minimal differences under left-handed circularly polarized (LCP) and right-handed circularly polarized (RCP) waves. The slight asymmetry observed in the total electric field amplitude suggests that the structure lacks significant structural chirality, a conclusion further supported by additional simulations.

\begin{figure*}[htbp]
  \centering
  \includegraphics[width=17.6cm]{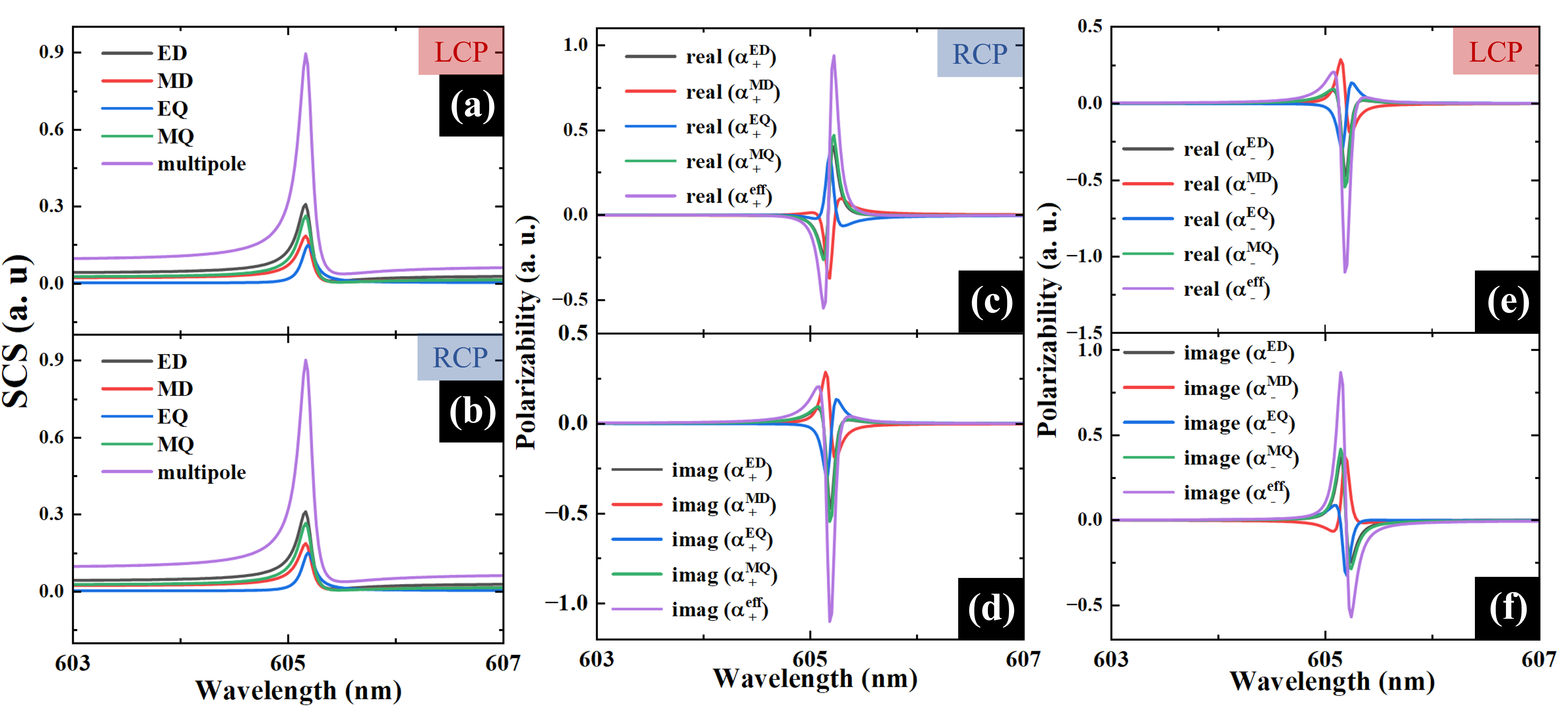}
  \caption{\textbf{Generalized chiral multipole mechanism applied in bilayer cylinder metasurface}.  (a) Generalized multipole scattering cross section under LCP waves. (b) Generalized multipole scattering under RCP waves. (c) The real part of each Generalized multipole effective polarizability under RCP waves. (d) The imaginary part of each Generalized multipole effective polarizability under RCP waves. (e) The real part of each Generalized multipole effective polarizability under LCP waves. (f) The imaginary part of each Generalized multipole effective polarizability under LCP waves. }
  \label{fig:S4}
\end{figure*}

The bilayer cylinder metasurface demonstrates a resonant point near 605 nm, as shown in Fig. \ref{fig:S2}(b). The reflection spectra under LCP and RCP waves are nearly identical, corroborating the findings from Fig. \ref{fig:S3}. The reflection and circular dichroism (CD) spectra calculated using our theoretical framework align well with the simulation results. Minor discrepancies in the CD spectra may stem from unaccounted multipole contributions, such as octupole moments. Furthermore, the scattering cross section is presented in Fig. \ref{fig:S3}(a) and (b). Under both LCP and RCP waves, the scattering cross section shows negligible differences, with the scattering fields predominantly governed by generalized electric dipole (ED) and magnetic quadrupole (MQ) modes. The effective polarizability for different generalized multipoles is depicted in Fig. \ref{fig:S4}(c)-(f). The effective polarizability, derived from the generalized chiral multipole mechanism, provides a more detailed insight into multipole interference compared to the scattering cross section.

In summary, while out-of-plane permittivity asymmetry persists and the structure exhibits variations in the real and imaginary parts of multipole moments under different circular polarizations, the overall behavior of multipole interference remains consistent. This indicates that the metasurface's response is primarily governed by multipole interactions rather than structural chirality.


\newpage
\section*{(\textbf{VI.}) Bilayer photonic crystal}
\begin{figure*}[htbp]
  \centering
  \includegraphics[width=17.6cm]{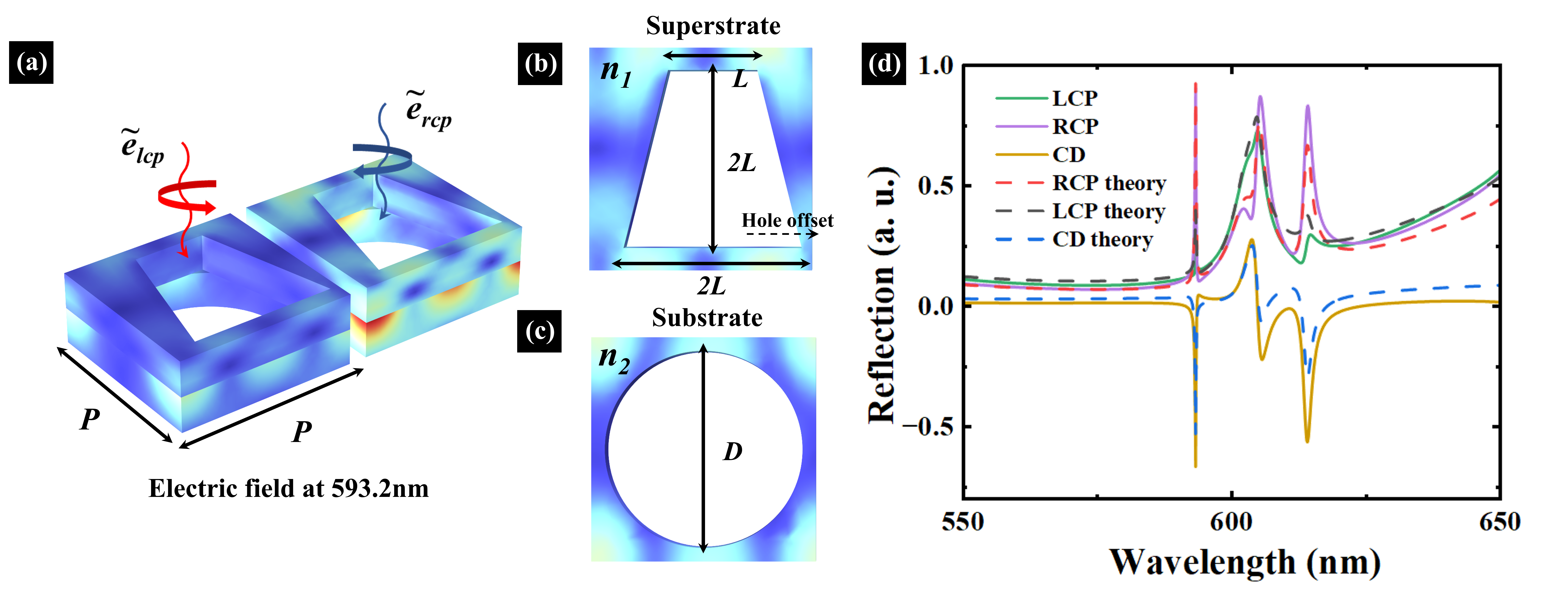}
  \caption{\textbf{Bilayer photonic crystal}. (a) The sketch of near fields (interms of total electric field amplitude $|\textbf{E}|$)at resonant point of bilayer cylinder under RCP and LCP waves. (b) Detail illustration for the superstrate. (c) Detail illustration for the substrate. (d) Reflection and CD spectra for simulation and theory.}
  \label{fig:S5}
\end{figure*}
\begin{figure*}[htbp]
  \centering
  \includegraphics[width=17.6cm]{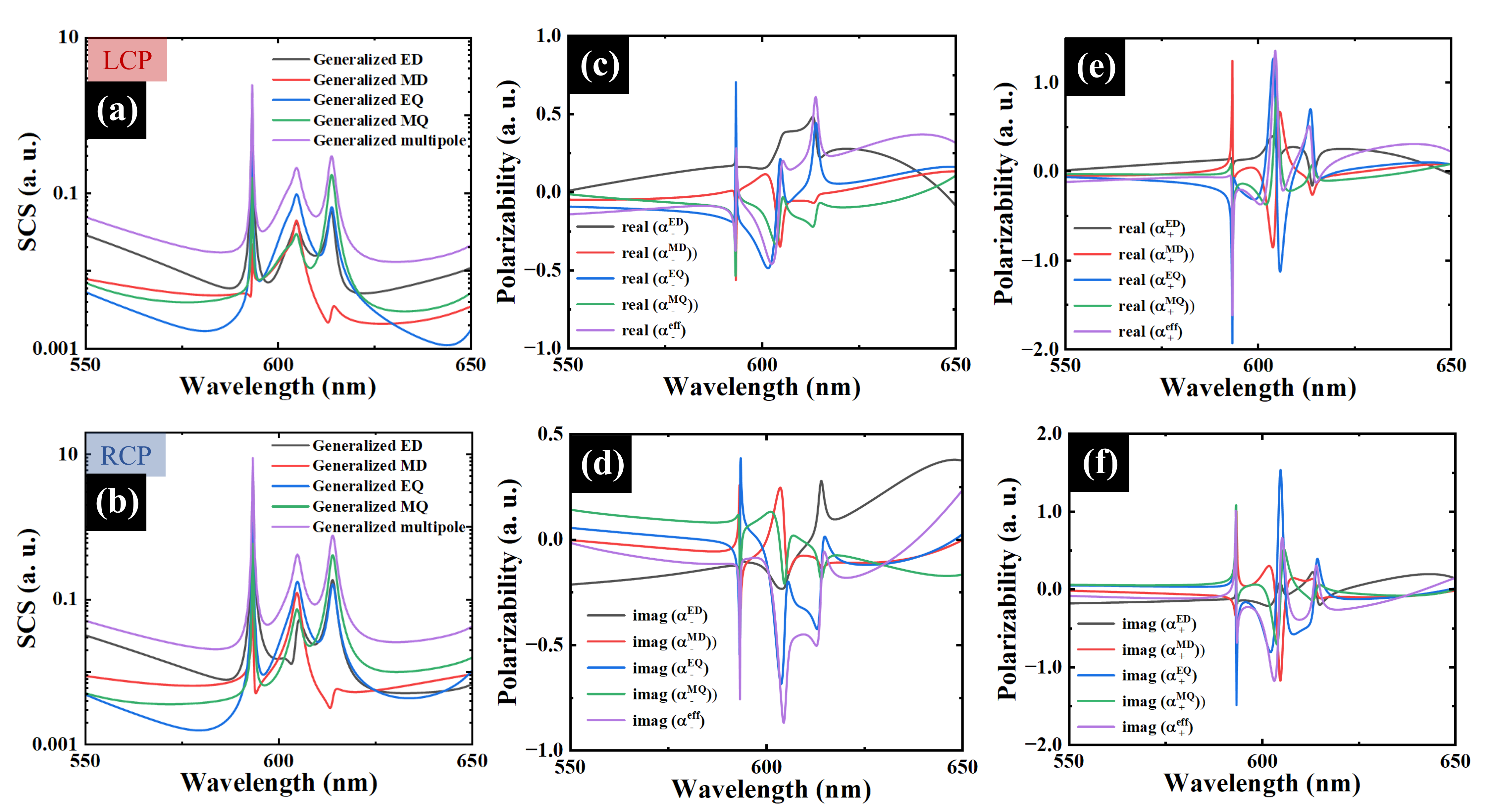}
  \caption{\textbf{Generalized chiral multipole mechanism applied in bilayer photonic crystal}.  (a) Generalized multipole scattering cross section under LCP waves. (b) Generalized multipole scattering under RCP waves. (c) The real part of each Generalized multipole effective polarizability under LCP waves. (d) The imaginary part of each Generalized multipole effective polarizability under LCP waves. (e) The real part of each Generalized multipole effective polarizability under RCP waves. (f) The imaginary part of each Generalized multipole effective polarizability under RCP waves. }
  \label{fig:S6}
\end{figure*}
In this section, we demonstrate one bilayer photonic crystal and validate our mechanism is still fit in this situation. As shown by Fig. \ref{fig:S5}(a), this nanostructures is composed of two layers photonic crystal. The superstate and substrate share the same period P of 500 nm and same thickness of 100 nm. Fig. \ref{fig:S5}(b) and (c) present the details of the upper layer and the down layer. A trapezoidal hole is dug in the upper layer of photonic crystal, and the central axis of the hole and the central axis of the period have an offset of 27 nm in the +x direction. The trapezoidal hole has an upper width L of 200nm and doubles when it comes to the lower width and the height. The substrate has a circle hole right in the central, whose diameter is 440 nm. The refractive index $n_1$ of superstate and $n_2$ of substrate is 2.5 and 1.5, seperately. The background refractive index is set to be 1.

Fig. \ref{fig:S5}(d) demonstrates the reflection and CD spectra for simulation and theory. It is clearly shown that the calculated reflection and CD spectra fit well with the simulation results.
Subsequently, the scattering cross section is presented in Fig. \ref{fig:S6} (a) and (b), while the effective polarizability for different generalized multipoles is illustrated in Fig. \ref{fig:S6} (c)-(f). By comparing the scattering cross section with the effective polarizability, it is evident that the Q-factor exhibits a strong correlation with the multipole moment. The Q-factor increases as the rate of change in the multipole moment becomes 
rapid. The resonant peaks observed in the reflection or transmission spectra are determined by multipole interference effects rather than being predominantly influenced by a single multipole moment.
  \end{document}